\documentclass[aps,pre,reprint,amsmath,amssymb,superscriptaddress,showkeys]{revtex4-1}

\usepackage[dvipsnames]{xcolor}
\usepackage[utf8]{inputenc}
\usepackage{amsmath, amsfonts, amsthm, bbm}
\usepackage{amsthm}
\usepackage{mathtools}
\usepackage{amssymb}
\usepackage{pifont}
\usepackage{bm}
\usepackage{graphicx} 
\usepackage{bmpsize}
\usepackage{tikz}
\usepackage{float}
\usepackage{braket}
\usepackage{hyperref}

\begin{document}
\title{Strategy Revision Phase with Payoff Threshold in the~Public~Goods~Game}
\author{Marco Alberto Javarone}
\email{marcojavarone@gmail.com}
\affiliation{Dipartimento di Fisica, Università di Bari, Bari, Italy}
\affiliation{Dutch Institute for Emergent Phenomena, Amsterdam, Netherlands}

\author{Shaurya Pratap Singh}
\affiliation{IISER Tirupati, Tirupati, Andhra Pradesh, India}
  
\begin{abstract}
Commonly, the strategy revision phase in evolutionary games relies on payoff comparison. Namely, agents compare their payoff with the opponent, assessing whether changing strategy can be potentially convenient.
Even tiny payoff differences can be crucial in this decision process.
In this work, we study the dynamics of cooperation in the Public Goods Game, introducing a threshold $\epsilon$ in the strategy revision phase. In doing so, payoff differences narrower than $\epsilon$ entail the decision process reduces to a coin flip.
Interestingly, with ordinary agents, results show that payoff thresholds curb the emergence of cooperation. Yet, the latter can be sustained by these thresholds if the population is composed of conformist agents, which replace the random-based revision with selecting the strategy of the majority.
To conclude, agents sensible only to consistent payoff differences may represent 'real-world' individuals unable to properly appreciate advantages or disadvantages when facing a dilemma. These agents may be detrimental to the emergence of cooperation or, on the contrary, supportive when endowed with a conformist attitude. 
\end{abstract}

\maketitle

\section{Introduction}
Understanding cooperation still represents an open challenge of paramount relevance~\cite{szolnoki00,duh01,perc00,xwang01}. From social scenarios to ecological systems, cooperation plays a non-trivial role in shaping the interactions among individuals, communities and groups.
These dynamics can be studied using the framework of Evolutionary Game Theory~\cite{perc01,santos01,perc03,javarone01,sanchez01} (EGT). The latter allows testing mechanisms devised for supporting the emergence of cooperation and modelling of real-world scenarios whose interactions rely on social dilemmas.
The simplest models relate to games with two strategies, i.e. usually cooperation and defection. More in detail, models represent a population whose agents interact through these games and undergo the following steps: (i) strategy selection, (ii) payoff collection, and (iii) strategy revision.
The latter, performed during the Strategy Revision Phase (SRP) process, is at the core of this investigation.
Before moving to details, let us remark on a growing literature presenting several ideas to sustain cooperation and explain its emergence.
To cite a few, in~\cite{danku01} authors proposed the presence of weak cooperators and weak defectors, in~\cite{nowak02,boyd01,helbing01,szolnoki02} authors studied the effect of punishing defectors, in~\cite{szolnoki03,dong01,javarone02} a conformist attitude revealed to sustain cooperation, in~\cite{moreno01} authors investigated costly-access environments, in~\cite{szolnoki07,feng01} authors analysed periodic payoff variations, then in~\cite{javarone03} payoff perturbations and risk perception (see also recent results on risk adversion~\cite{armas01}) resulted a beneficial element for supporting cooperative behaviours. We end this very brief list by mentioning relevant works based on complex interaction topologies (i.e. structures for connecting agents)~\cite{szabo01,santos02,moreno02,santos03,arenas01,szolnoki01,nowak03}, which are known to be fundamental in driving the population towards cooperation. Most of these mentioned studies build on Nowak's seminal work~\cite{nowak01} about the five rules for the evolution of cooperation, proposing methods to exploit rules such as direct reciprocity, kin selection, and so on.\\
Going back to our investigation, we analyse an aspect related to the SRP considering a usual mechanism requiring agents to compute the payoff difference, i.e. compare their payoff with that of an opponent to take a decision.
Commonly, the payoff difference, no matter its value, gets enhanced by the (inverse of) system temperature. Here, we add a step introducing a threshold. The latter applies to the payoff difference so that differences smaller than the threshold are set to zero, leading agents to ignore them. \\
Accordingly, agents alternate rational thinking (i.e. purely driven by payoff differences) with a coin flip-based method, which typically applies only when their payoff is the same as the payoff of their opponents.\\
Therefore, the proposed model represents real-world social scenarios whose individuals may be sensitive only to consistent payoff differences or be unable to appreciate small (dis/)advantages.
Then, given the relevance of conformism in several social scenarios~\cite{sznajdweron01,sznajdweron02}, we include this behaviour in our study.
Eventually, among the games we can consider for analysing the proposed model, we choose the Public Goods Game (PGG)~\cite{perc04,perc05,han01,wang01,amaral01}, also known as $n$-person Prisoner's Dilemma.
Interestingly, results achieved through numerical simulations show that neglecting even tiny payoff differences can affect the usual evolution of cooperation in the PGG. Also, the main difference relates to outcomes obtained with game settings that sustain strategy co-existence.
The remainder of this manuscript is organised as follows. Section~\ref{sec:model} introduces the proposed model, Section~\ref{sec:results} shows the results of numerical simulations and finally, Section~\ref{sec:conclusion} discusses the main findings and concludes by mentioning possible future developments.
\section{Model}\label{sec:model}
Let us begin by briefly describing the PGG and the mechanism we introduce for studying the potential effect of a threshold for the payoff difference.
The PGG considers a population of $N$ agents, here arranged over a regular square lattice with continuous boundary conditions. Agents can contribute to their communities of belonging or behave as free riders. Namely, they can act as cooperators or defectors, respectively.
Communities are identified through direct interactions, while contributions are mapped to a token whose collection represents the 'public goods'. 
Then, the latter is enhanced by a synergy factor $r$ representing the parameter the system can exploit to support, or less, cooperative behaviours.
Eventually, the public goods accumulated in each community are divided among all the agents, no matter their strategy. 
Since agents do not know the opponents' strategy before acting, defection is much less risky than cooperation.
In a square lattice, communities have a size $G = 5$, and each agent belongs to $5$ different communities identified through direct connections. For instance, an agent identifies its core community by considering itself and its four neighbours.\\
This game has the following payoff structure:
\begin{equation}\label{eq:payoff}
\left\{\begin{array}{l}\pi^{c}=\frac{rN^{c}}{G}-c \\ \pi^{d}=\frac{r N^{d}}{G}\end{array}\right.      
\end{equation}
\noindent where $\pi^{d}$ and $\pi^{c}$ denote the payoff of defectors and cooperators, respectively, while $N^c$ and $N^d$ are the number of cooperators and defectors within a group, respectively. Then, without loss of generality, we set to $1$ the value of the token $c$.
After each round of the game, agents can change their strategy by undergoing the Strategy Revision Phase that, as previously mentioned, can be implemented according to various mechanisms.\\
Here, we consider a stochastic SRP based on the following Fermi-like rule:
\begin{equation}\label{eq:fermi}
W(s_{x} \leftarrow s_{y}) = \left(1+\exp \left[\frac{\pi_{x}-\pi_{y}}{K}\right]\right)^{-1}
\end{equation}
\noindent where $T$ denotes the system temperature, $\pi_y$ the opponent's payoff, and $\pi_x$ the payoff of the agent undergoing the strategy revision. So, equation~\eqref{eq:fermi} corresponds to the probability the $x$-th agent selects the opponent's strategy for the next round.\\ 
The temperature parameter, also known as noise, allows tuning the degree of rationality of agents~\cite{javarone04}. Namely, low temperatures entail agents acting rationally, while high temperatures make the SRP equivalent to a coin flip (i.e. a random process) so that agents behave irrationally.
In summary, a time step includes the following actions: agents play the PGG in their groups of belonging, accumulate a payoff, and, if at least one of their direct neighbours has a different strategy, undergo the SRP.
The above dynamics are repeated till an equilibrium of order, or a steady state, is reached. The order emerges if one strategy prevails, while a steady state represents a stable strategy co-existence.
\subsection*{Payoff Difference Threshold}
Now, we focus on the payoff difference agents compute in the SRP, i.e. the term $\Delta \pi = \pi_x - \pi_y$ ---see Eq.~\eqref{eq:fermi}. 
Introducing a threshold $\epsilon$ to consider only relevant $\Delta \pi$ values entails adding a simple step to the above-described dynamics. Thus, once defined $\epsilon$, all $\Delta \pi$ smaller than this threshold are set to zero. For the sake of clarity, we consider the absolute value, i.e. $|\Delta \pi| < \epsilon \to \Delta \pi = 0$, so the agent performing the comparison neglects both advantages and disadvantages if not large enough.\\
According to the proposed model, the relevant variables are the synergy factor $r$ and the $\epsilon$ threshold. 
Notice that the latter has to be defined for each specific value of the former. Also, we used the following methods to compute $\epsilon$:
\begin{itemize}
    \item $\epsilon = \langle\Delta \pi\rangle$;
    \item $\epsilon = \langle\Delta \pi\rangle + \sigma$
\end{itemize}
\noindent where $\langle\Delta \pi\rangle$ denotes the average value of the payoff difference measured in the classical setting of the PGG during the first $1000$ time steps. More specifically, at each time step, we compute the average payoff difference that agents report with their opponents, and then we calculate the average along all time steps.
Similarly, we compute the average standard deviation $\sigma$, in the second definition of $\epsilon$.\\
Eventually, we study the proposed model on two populations, i.e. composed of ordinary agents and conformist agents. 
Ordinary agents choose their strategy randomly when $\Delta \pi < \epsilon$. On the other hand, conformist agents are those that, in the same condition (i.e. $\Delta \pi < \epsilon$), prefer to choose the strategy of the majority in their core group (i.e. the group containing all four nearest neighbours) ---see also~\cite{galam02,galam03,crokidiakis01}.
\section{Results}\label{sec:results}
Numerical simulations, performed with a population of $N = 10000$ agents arranged over a regular square lattice with continuous boundary conditions at a temperature $T = 0.5$, consider an initial density of cooperators $\rho_c = 0.5$ (randomly distributed) ---see also~\cite{galam01}.
\begin{figure*}[ht!]
    \includegraphics[scale=0.25]{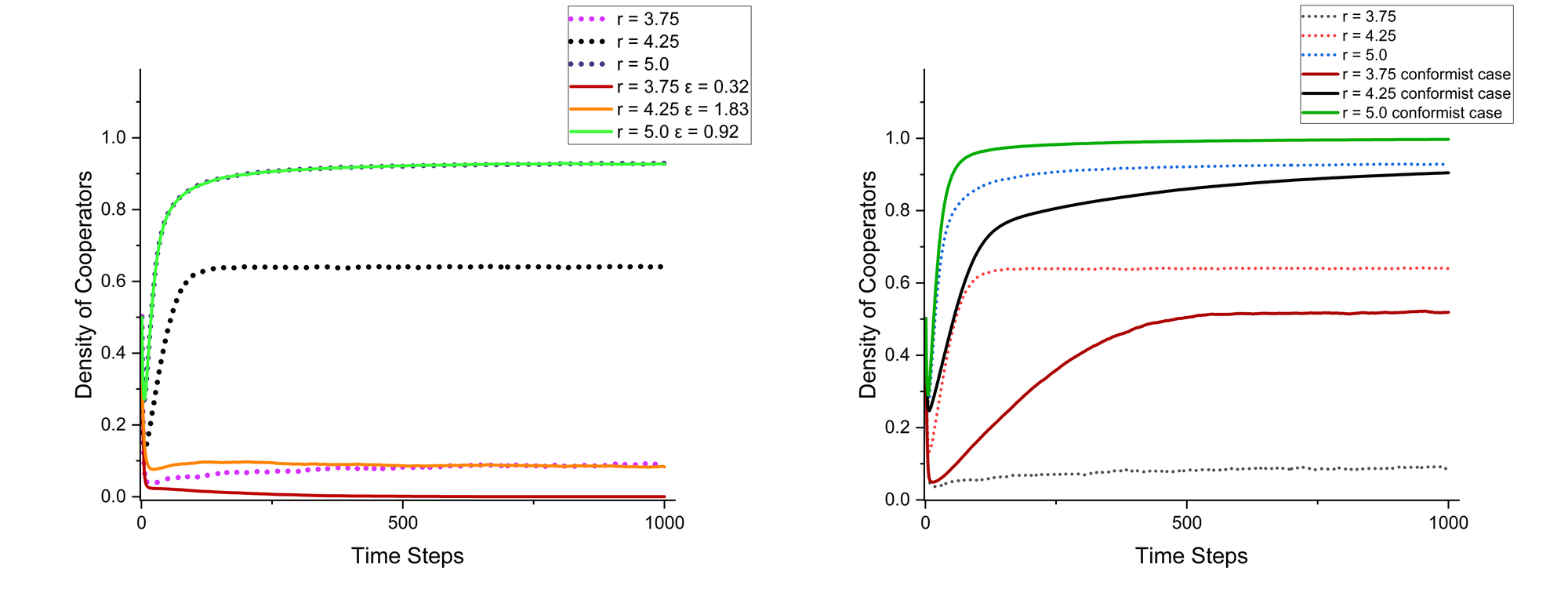}
    \caption{The density of cooperators achieved by $\epsilon = \langle \pi \rangle$, with three different synergy factors: $r = 3.75$ ($\epsilon = 0.32$), $r = 4.25$ ($\epsilon = 1.83$), and $r = 5.00$ ($\epsilon = 0.92$) (see the legend). On the left, ordinary agents and, on the right, conformist agents. Results are averaged over $100$ simulation runs.}\label{fig:figure_1}
\end{figure*}
In the described conditions, there is a critical threshold for the synergy factor $r_{th} = 3.72$ (e.g. see~\cite{perc04}), such that values $r < r_{th}$ lead the population to converge towards an equilibrium of full defection. On the other hand, values of the synergy factor in the range $r_{th} < r < 5$ lead to a steady state characterised by the co-existence of strategies. Eventually, for values $r > 5$, cooperation becomes the dominant strategy, leading the population towards an ordered equilibrium.
Given this premise, our first task is computing the $\epsilon$ value on varying the synergy factor $r$.
To this end, for each specific $r$, we perform $100$ numerical simulations with the above-described setting (i.e. number of agents, temperature, etc). 
\begin{figure*}[ht!]
    \includegraphics[scale=0.25]{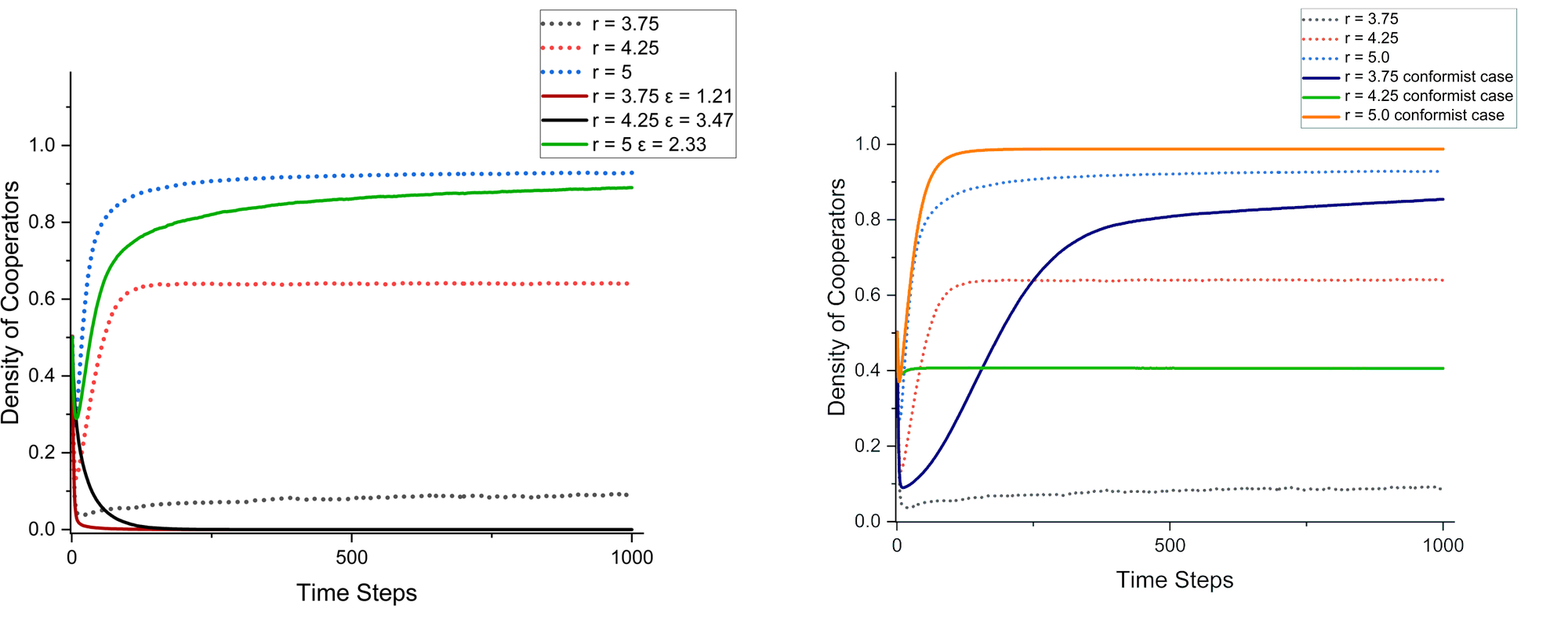}
\caption{The density of cooperators achieved by $\epsilon = \langle \pi \rangle + \sigma$, with three different synergy factors: $r = 3.75$ ($\epsilon = 1.21$), $r = 4.25$ ($\epsilon = 3.47$), and $r = 5.00$ ($\epsilon = 2.33$) (see the legend). On the left, ordinary agents and, on the right, conformist agents. Results are averaged over $100$ simulation runs.}\label{fig:figure_2} 
\end{figure*}
\begin{figure*}
\includegraphics[scale=0.25]{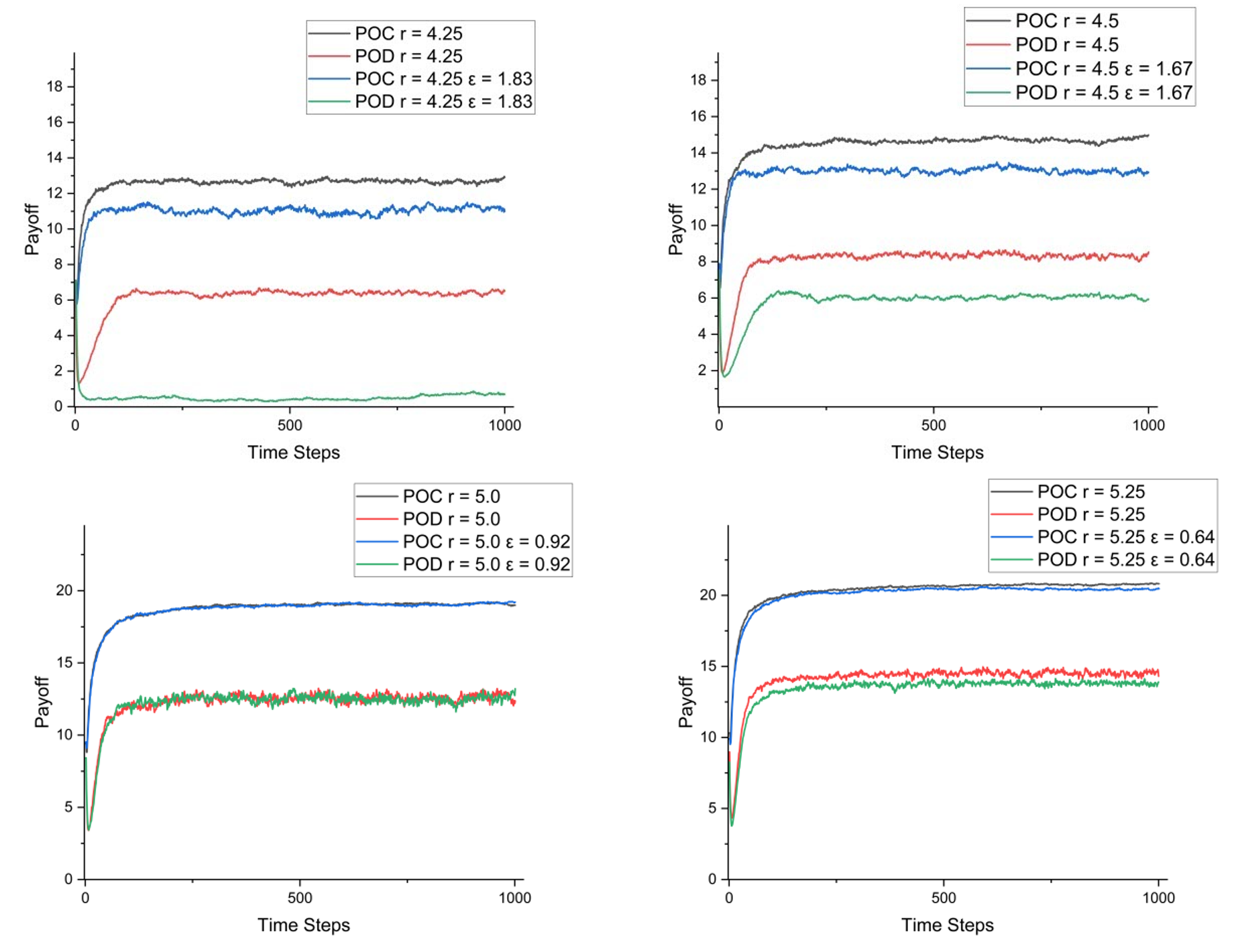}
\caption{The payoff of cooperators and defectors for three different synergy factors: $r = 4.25$ (top left) with $\epsilon = 1.83$, $r = 4.5$ (top right)  with $\epsilon = 1.67$, $r = 5.0$ (bottom left) with $\epsilon = 0.92$, and $r = 5.25$ (bottom right) with $\epsilon = 0.64$, achieved with ordinary agents. The higher the synergy factor, the smaller the difference between the classical PGG and the proposed model. Results are averaged over $10$ simulation runs.}\label{fig:figure_3}
\end{figure*}
\begin{figure*}
\includegraphics[scale=0.25]{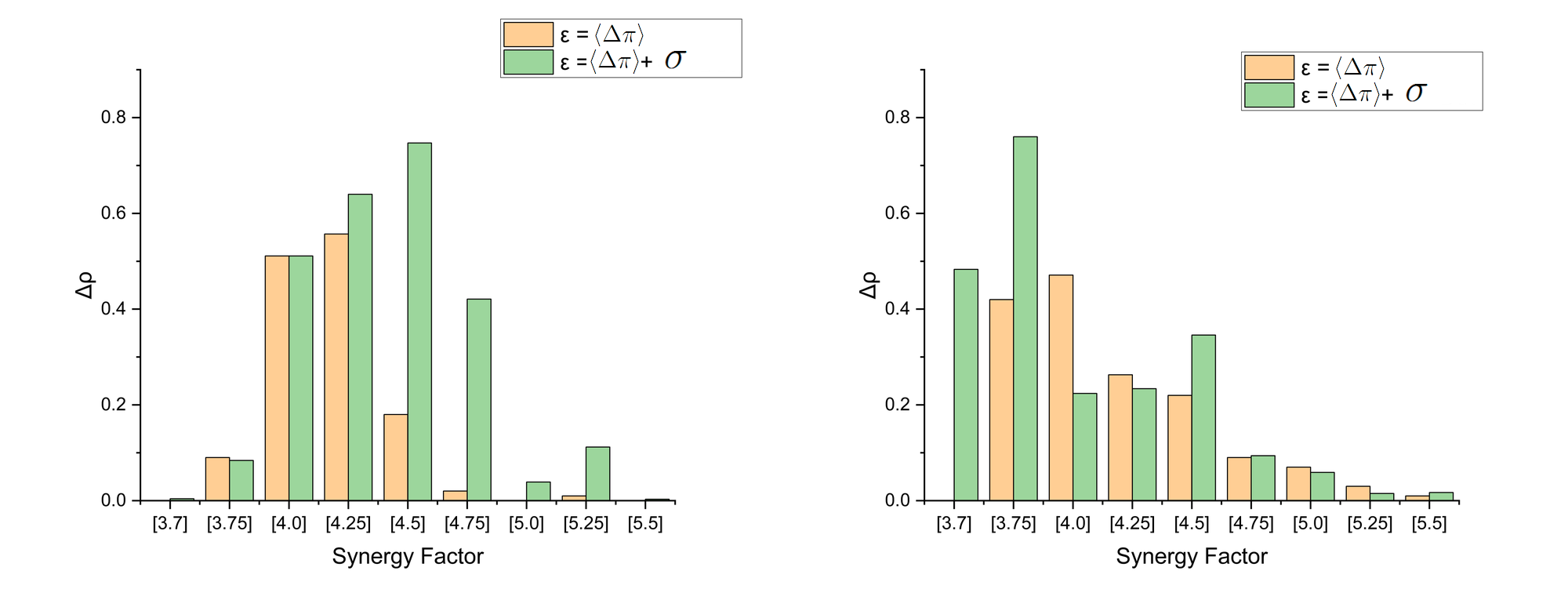}
\caption{The difference in density of cooperators $\Delta \rho$ achieved in ordinary populations (left-hand) and conformist populations (right-hand), compared to the classical PGG, on varying the synergy factor $r$. As indicated in the legend, for each population, we include both $\epsilon$ values (see main text for the definitions). Results are averaged over $100$ simulation runs.}\label{fig:figure_4} 
\end{figure*}
Table~\ref{tab:epsilon_table} shows the results of this preliminary analysis.
\begin{table}[h!]
    \centering
    \begin{tabular}{|c|c|c|}
    \hline
         Synergy Factor $r$ & $\epsilon \coloneqq \langle \pi \rangle$ & $\epsilon \coloneqq \langle \pi \rangle + \sigma$ \\
         \hline
         3.7& 0.07 & 0.42\\
         \hline
         3.75& 0.32 & 1.21\\
         \hline
         4.0& 1.74 &  3.3\\
         \hline
         4.25& 1.83  & 3.47\\
         \hline
         4.5& 1.67 & 3.34\\
         \hline
         4.75& 1.33 & 2.93\\
         \hline
         5.0& 0.92& 2.33\\
         \hline
         5.25& 0.64 & 1.82\\
         \hline
         5.5& 0.45 & 1.42\\
         \hline
    \end{tabular}
    \caption{Synergy factor $r$ and threshold $\epsilon$ defined according to the two presented methods (see text). Results are averaged over $100$ simulation runs.}
    \label{tab:epsilon_table}
\end{table}
So, now we have all the elements to implement the main simulations. 
For clarity, we study the dynamics of the PGG by implementing an SRP based on Eq~\eqref{eq:fermi}, introducing as an additional step the comparison between the $|\Delta \pi|$ term and the $\epsilon$ threshold.
We recall that $\Delta \pi$ appears in explicit form in Eq~\eqref{eq:fermi}, i.e. $\pi_x -\pi_y$, and here we consider its absolute value.
Thus, using the $\epsilon$ values shown in Table~\ref{tab:epsilon_table}, if $|\Delta \pi| < \epsilon$, the payoff difference is set to zero.
Notice that the above rule applies only to ordinary agents since conformist agents choose the strategy of the majority in their neighbourhood if $|\Delta \pi|$ is below the threshold.

Following the same method implemented for computing the $\epsilon$ values, we measure the fraction of cooperators and the payoff trends resulting from the two strategies in different conditions.
Figure~\ref{fig:figure_1} shows the result of the first comparison between the density of cooperators in the two populations, i.e. ordinary agents and conformist agents, on varying the synergy factor, using the threshold $\epsilon$ defined according to the average payoff (i.e. first definition).
In populations composed of ordinary agents, using low and intermediate synergy factors, the threshold $\epsilon$ curbs cooperation. In addition, its effect vanishes as $r$ gets close to $5$ (or higher).
On the other hand, populations composed of conformist agents reach a higher density of cooperation for all considered synergy factors. Notice that we cannot compare this specific population with another one that includes conformists and does not use a payoff threshold. Regarding this, we recall that here the conformist behaviour emerges only when $|\Delta \pi| < \epsilon$. 
Since conformism supports cooperation for all values of $r$, our results confirm its beneficial effect for this strategy, as already observed in previous studies on dilemma games~\cite{dong01}.
Then, the same analysis is performed by considering the second definition of $\epsilon$, i.e. the one including the average standard deviation of the payoff difference ---see Figure~\ref{fig:figure_2}. 
In general, results show that the higher the $\epsilon$, the smaller the final density of cooperators. Yet, beyond observing that conformist agents are more likely to cooperate, we find an interesting phenomenon in this population.
Namely, intermediate values of synergy factors enhance cooperation less than small values (all higher than $r_{th}$). That is due to the combination of large $\epsilon$ values (for those synergy factors) and the uncertainty about the considered social behaviour in supporting cooperation in these conditions. 
In a few words, in most cases, conformity clearly supports cooperation, albeit with high thresholds its beneficial effect is reduced.
Before moving further, let us emphasise that in ordinary populations, the higher the threshold, the higher the difference in the final density of cooperators between the classical PGG and the proposed model.
We can now focus on another parameter of interest, i.e. the payoff trend of the two strategies measured during the system evolution. Looking at the results in Figure~\ref{fig:figure_3}, the higher the synergy factor, the smaller the difference between the proposed model and the classical PGG. The difference between the payoff trends for both strategies becomes larger by reducing the value of $r$.
Finally, we analyse the difference in the density of cooperators $\Delta \rho$, achieved in each considered condition, on varying the synergy factor with the normal PGG dynamics. 
This comparison includes both populations, i.e. ordinary and conformist agents ---see Figure~\ref{fig:figure_4}.
In the population composed of ordinary agents, the highest differences are obtained for intermediate synergy factors, i.e. $4 \leq r \leq 4.5$. Yet, a clear peak at $r = 4.5$ is detected using the threshold defined through the average and the standard deviation of the payoff differences (i.e. the second definition of $\epsilon$).
In the conformist population, the distribution of $\Delta \rho$ is different as the peaks are obtained for low values of synergy factors, and $\Delta \rho$ decreases as $r$ becomes larger than $4.5$.
In both populations, synergy factors higher than $5$ tend to make poorly relevant the effect of $\epsilon$ in the SRP.
\section{Discussion and Conclusion}\label{sec:conclusion}
In this work, we analysed the strategy profile of a population interacting through an evolutionary game whose agents are, by design, unable to give value to payoff differences below a given threshold. 
To this end, we focused on the PGG, which is particularly useful for studying social dilemmas (as well as other scenarios, e.g.~\cite{kurokawa01,liu01}).
The proposed model introduces a tiny variation to the classical setting of this game. More specifically, starting with a population arranged over a square lattice and at a low temperature, we enriched the strategy revision phase, adding a threshold to the payoff difference. 
In doing so, payoff differences which do not satisfy the introduced constraint are set to zero, making the SRP equivalent to a coin flip.
Then, given the high relevance of conformism in supporting cooperation~\cite{szolnoki03,javarone02}, we considered ordinary and conformist agents separately. In both cases, the above-mentioned low temperature ensures agents behave rationally when the payoff difference is higher than/equal to the threshold. On the other hand, in the opposite event, i.e. with a payoff difference tinier than the threshold, conformist agents select the strategy of the majority holding in their core group, while ordinary agents, as described, take a coin flip-based decision.
Results of numerical simulations suggest that introducing a threshold in the SRP with ordinary agents may curb cooperation. More in detail, if agents ignore even 'tiny' payoff differences, cooperation survives/prevails when the synergy factor is big enough.
On the other hand, as mentioned, conformity plays a fundamental role in curbing defection. So, although conformism relates to a fraction of SRP processes, we confirmed its relevance in these dynamics.

To conclude, results suggest that individuals unable to give value to limited payoff differences can be detrimental to cooperation. Yet, this attitude, with conformist behaviour, can lead to the opposite scenario where cooperation is supported. That leads us to further observations.
First, in social contexts, stimulating cooperation is far from trivial. So, the presence of individuals behaving similarly to our ordinary agents may constitute one of the possible motivations. 
Yet, mixing conformity and rationality resulted in a highly beneficial support for cooperation. Regarding this, we find reasonable individuals may behave extremely rationally when observing their payoff is much higher/lower than that collected by an opponent and imitate their peers when such difference is less evident.
In summary, the ability to compute and appreciate even tiny advantages and disadvantages can be fundamental for the evolution of cooperation in dilemma games.
To conclude, further investigations into this direction may concern other social dilemmas, more complex population structures (e.g. scale-free networks~\cite{estrada01}), additional SRP mechanisms, more social behaviours such as anti-conformity~\cite{calvelli01,krueger01}, and the application of the proposed model beyond social scenarios, e.g. ecological systems~\cite{javarone05}, economic systems~\cite{javarone06,wu01}, and many others.

\acknowledgments
MAJ is supported by the PNRR NQSTI (Code: $PE23$). 

\end{document}